\documentclass[pdflatex,sn-mathphys,Numbered]{sn-jnl}

\usepackage{graphicx}%
\usepackage{multirow}%
\usepackage{amsmath,amssymb,amsfonts}%
\usepackage{amsthm}%
\usepackage{mathrsfs}%
\usepackage[title]{appendix}%
\usepackage{xcolor}%
\usepackage{textcomp}%
\usepackage{manyfoot}%
\usepackage{booktabs}%
\usepackage{graphicx}

\theoremstyle{thmstyleone}%
\theoremstyle{thmstyletwo}%
\theoremstyle{thmstylethree}%

\begin{document}

\title[Classical and Hybrid Quantum Learning for Trigger-Like Event Selection]{Classical and Hybrid Quantum Machine Learning for Trigger-Like Event Selection on CMS Open Data: An Eight-Qubit, PCA-Constrained Benchmark}

\author*[1]{\fnm{Tariq} \sur{Mahmood}}\email{tariqmahmood.chep@pu.edu.pk}
\author[1]{\fnm{Muhammad Awais} \sur{Rafique}}
\author[1]{\fnm{Talab} \sur{Hussain}}
\author[2]{\fnm{Juan Pablo} \sur{P\'erez Aguilar}}
\author[3,4]{\fnm{Alfredo} \sur{Raya}}
\author[1]{\fnm{Muhammad} \sur{Ahsan}}

\affil[1]{\orgdiv{Centre for High Energy Physics}, \orgname{University of the Punjab, Lahore}, \orgaddress{\country{Pakistan}}}
\affil[2]{\orgdiv{Facultad de Ciencias F\'isico-Matem\'aticas}, \orgname{Universidad Michoacana de San Nicol\'as de Hidalgo}, \orgaddress{\city{Morelia}, \state{Michoac\'an}, \country{Mexico}}}
\affil[3]{\orgdiv{Facultad de Ingenier\'ia El\'ectrica}, \orgname{Universidad Michoacana de San Nicol\'as de Hidalgo}, \orgaddress{\city{Morelia}, \state{Michoac\'an}, \country{Mexico}}}
\affil[4]{\orgdiv{Centro de Ciencias Exactas}, \orgname{Universidad del B\'io-B\'io}, \city{Chill\'an}, \country{Chile}}

\abstract{Event triggering sits at the heart of high-energy physics, where the rare events of interest must be retained while an overwhelming background is discarded under tight latency and bandwidth budgets. This work compares four classical machine-learning models, namely a support vector machine, an artificial neural network, a convolutional network and a long short-term memory network, with four hybrid quantum counterparts, on a trigger-like binary classification task built from CMS open data. The label is defined by an invariant-mass window, and the inputs combine reconstructed kinematics with physics-motivated derived variables: the pseudorapidity difference, the wrapped azimuthal difference, the angular separation and the total transverse momentum. The quantum models run under a fixed resource budget of eight qubits, a principal-component compression to sixteen features and state-vector simulation. Every model shares the same stratified split, the same preprocessing and a common decision threshold, and performance is reported through accuracy, ROC-AUC, F1-score, precision and recall. The strongest classical model is the artificial neural network, at 93.53 percent accuracy and 0.9819 ROC-AUC, while the strongest quantum model is the quantum convolutional network, at 90.89 percent accuracy and 0.9731 ROC-AUC, with the quantum neural network close behind. The quantum-kernel and recurrent-quantum approaches trail both, which places the trainable hybrid embeddings ahead within this budget. The study is meant as a controlled reference point rather than a claim of quantum advantage.}


\maketitle

\section{Introduction}\label{sec:intro}

Particle physics studies the smallest constituents of matter and the carriers of the fundamental forces, together with the way those constituents interact. Its objects carry attributes such as electric charge, spin and mass, and although each object is complex, it is convenient to treat it as point-like. The theoretical language for all of this is quantum field theory, in which symmetry plays a guiding role \cite{britannica_particle}. The Standard Model brings two such theories together: the electroweak theory, which accounts for the electromagnetic and weak forces, and quantum chromodynamics, which accounts for the strong force. Both are gauge theories, so interactions are described through the exchange of messenger particles that carry one unit of spin \cite{britannica_sm}. Everything the Standard Model predicts about quarks, leptons and bosons has been probed by an experimental program whose largest instrument is the Large Hadron Collider \cite{fsu_hep,iccub}.

Experimental high-energy physics produces data at a rate that no downstream analysis could ever store in full. A trigger is the online decision system that inspects each collision and keeps only the events worth recording. Because the interesting processes are rare and the background is enormous, the trigger has to be both selective and fast, and it operates under strict limits on latency and on the bandwidth available for readout. This is the setting that motivates the present study: it treats event selection as a supervised classification problem, in which each event is mapped to a label that marks it as signal or background \cite{luongo_ml}.

Machine learning offers a natural toolbox for that mapping. In the supervised setting, every example carries a known target, and when the target is a category the task is classification. Over the past decade, deep networks have become the standard way to learn such mappings from collider data. Quantum machine learning proposes a different route, in which quantum circuits encode the data and a measurement returns a score. The appeal is that superposition and entanglement give access to a feature space that is expensive to reach classically \cite{csiro_qml,dunjko_review}. Quantum mechanics supplies the physical principles behind that idea \cite{sciencedirect_qm}, and quantum computing supplies the hardware and the algorithms, still at an early stage but advancing quickly \cite{ibm_qc}.

The question this work addresses is a practical one. When the same trigger-like task, the same data and the same evaluation protocol are handed to classical and to hybrid quantum models, and when the quantum models are held to a realistic resource budget, how do the two families compare? To answer it, we build a binary classification task from CMS open data, define the label through an invariant-mass window, describe each event with reconstructed kinematics and a handful of physics-motivated derived variables, and train eight models under a shared protocol. Four are classical and four are hybrid quantum. The quantum models are restricted to eight qubits, to a principal-component compression of the inputs to sixteen features and to state-vector simulation, so that the comparison reflects the constraints under which such models are realistically trained today.

\section{Related work}\label{sec:related}

A first line of work brings classical deep learning to online selection and to event classification. A quantum-inspired tree tensor network trained on LHCb data reaches about 70.5 percent accuracy on real-time b-jet tagging, matching a deep network and improving on conventional muon tagging \cite{felser_ttn}. Generative models have been adopted to encode collision data with physics constraints so that collider simulation becomes faster and more reproducible \cite{butter_lhc}. In sPHENIX, where the 15 kHz readout limits the reach for heavy-flavor events, a real-time trigger built from multilayer perceptrons and graph networks selects such events with an efficiency near 0.945 at a useful purity \cite{chen_sphenix}. Field-programmable gate arrays have been used to run inference for long-lived decays at the level of milliseconds per event, with an autoencoder reaching roughly 3.1 to 3.7 ms and 553 frames per second on one board and a convolutional network reaching 2.6 ms and 1497 frames per second on another, within second-level trigger requirements \cite{coccaro_fpga}. A bonsai boosted decision tree, whose splits are pruned to fixed discrete values, avoids the slowness of unpruned classifiers; it is somewhat less efficient than a standard tree on four-body signals but improves on five-body signals and greatly reduces instability \cite{gligorov_bbdt}. A graph-network approach to full-event interpretation reconstructs heavy-hadron decay chains after the LHCb upgrade, capturing about 94 percent of the true b hadrons while removing 96 percent of the background and reducing the number of particles per event by more than an order of magnitude \cite{pardinas_dfei}. End-to-end convolutional models trained on high-fidelity CMS calorimeter images learn discriminative features directly, reaching a ROC-AUC of 0.807 for shower classification and separating electron-positron pairs from photon pairs at a ROC-AUC near 0.997 \cite{andrews_e2e}. Deep networks with spatial and temporal labels have also been used to forecast extreme events in nonlinear systems, where a residual network reaches an accuracy of about 91 percent \cite{jiang_extreme}, and Lorentz-equivariant networks trained on raw four-vectors improve quark-gluon separation under realistic conditions \cite{kasieczka_qg}.

A second line explores quantum kernels. A quantum support vector machine, evaluated on up to twenty qubits and fifty thousand events across simulators and IBM hardware, rejects roughly 92 percent of the background at 15 qubits while retaining 70 percent of the signal, and its performance stays close to optimal as the qubit count grows from ten to twenty \cite{wu_qsvm}. On continuum suppression in B-meson decays, physics-inspired encodings raise the best area under the curve to 0.848, against 0.793 for a plain support vector machine, though the best result on real hardware settles near 0.703 \cite{heredge_qsvm}.

A third line studies variational quantum classifiers. A ten-qubit variational classifier separates rare Higgs events from background with classifier scores in the range 0.81 to 0.83 on simulators and 0.81 to 0.82 on hardware, matching classical support vector machines and boosted trees and improving the signal-to-background ratio \cite{wu_qvc}. A two-qubit hybrid classifier trained with quantum gradient descent converges faster and reaches an area under the curve of 0.794, above a standard variational classifier at 0.773 and a classical network at 0.738 \cite{blance_vqc}. Broader comparisons of quantum circuit learning and variational classifiers report areas under the curve between 0.80 and 0.85, on par with boosted trees and deep networks, with stable training and convergence on real machines \cite{terashi_qml}. Quantum convolutional networks trained on DUNE data match earlier networks at equal parameter count and raise test accuracy for muon-proton discrimination \cite{chen_qcnn}, and a study of vector boson scattering finds that variational circuits reach areas under the curve comparable to deep networks, occasionally higher, with few input variables and few training events \cite{cugini_vbs}. Surveys of the field are careful to note that quantum machine learning remains exploratory, with noisy hardware and small datasets that still limit its reach \cite{guan_qml}.

A fourth line pursues model-independent discovery. Variational autoencoders trained only on Standard Model events flag anomalous collisions without assuming a specific new-physics model, selecting rare events at a rate near $5.4\times 10^{-6}$ of the traffic while preserving strong separation across a range of scenarios \cite{cerri_vae}. Foundation models add a complementary angle: a graph network pretrained on 120 million events across twelve processes learns transferable representations, improves accuracy by more than five percent in the low-data regime and reaches high representational fidelity at a fraction of the training cost \cite{ho_pretrained}.

The present study sits alongside these efforts but differs in emphasis. Rather than optimizing a single model, it fixes a task, a dataset and a protocol, and then compares a classical and a hybrid quantum family under an explicit and deliberately modest quantum budget. The intent is a controlled reference point.

\section{Data and task definition}\label{sec:data}

\begin{figure}[h]
\centering
\includegraphics[width=0.72\textwidth]{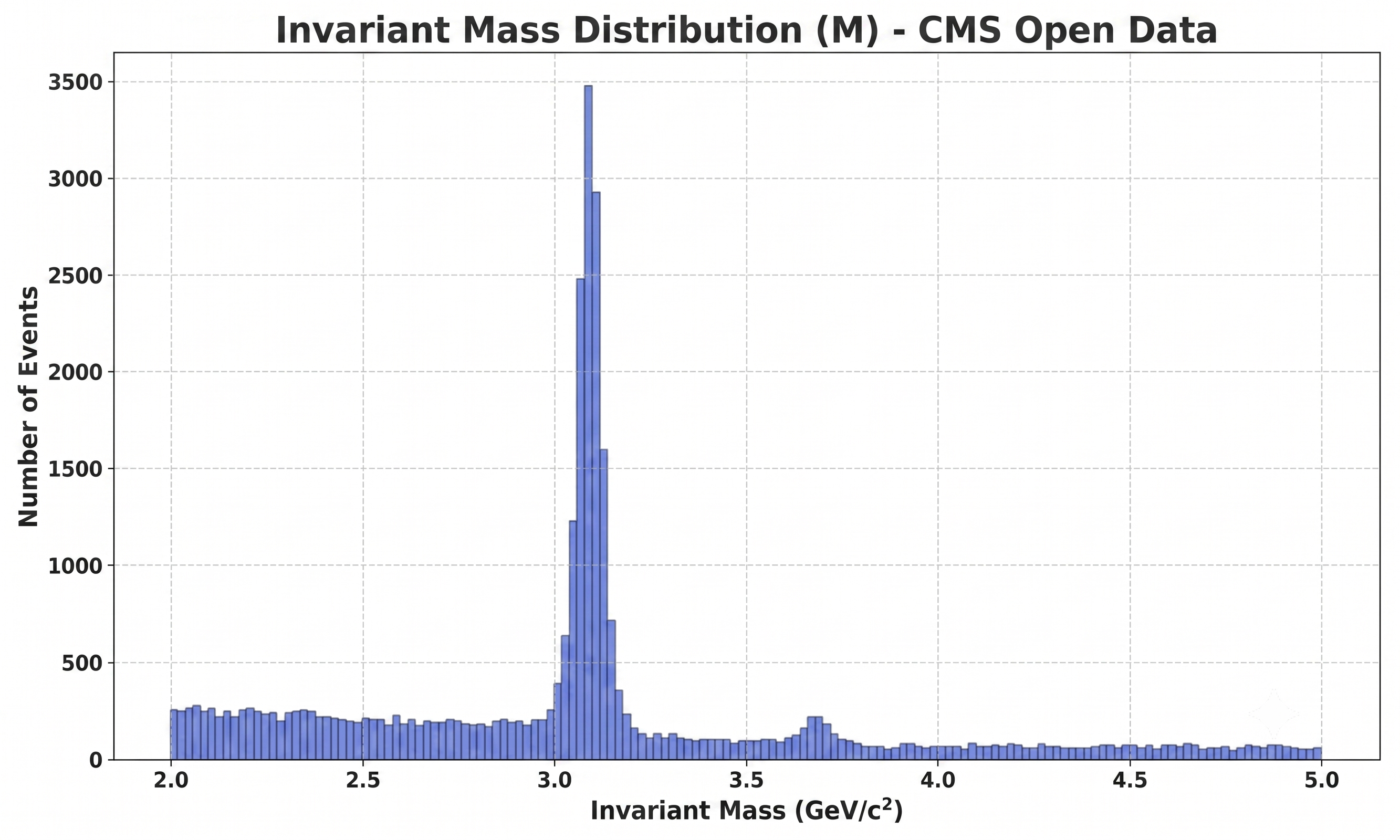}

\caption{Invariant-mass distribution of the CMS sample. The shaded band marks the trigger window $3.0 \le M \le 3.2$ GeV that defines the positive label.}\label{fig:massdist}
\end{figure}

The task is a binary trigger-like classification built from CMS open data \cite{cms_data}, comprising 31{,}892 events and 21 columns. For each event the measured four-vector components of two reconstructed objects are combined into the di-object invariant mass,
\begin{equation}
M = \sqrt{(E_1+E_2)^2 - (p_{x1}+p_{x2})^2 - (p_{y1}+p_{y2})^2 - (p_{z1}+p_{z2})^2}.\label{eq:mass}
\end{equation}
The label follows a window on this quantity: an event receives $y=1$ when $3.0 \le M \le 3.2$ GeV, the region of the $J/\psi$ resonance, and $y=0$ otherwise. The distribution of $M$ and the selected window are shown in Fig.~\ref{fig:massdist}. The raw kinematic variables are supplemented by derived quantities that are standard in the field: the pseudorapidity difference $\Delta\eta$, the wrapped azimuthal difference $\Delta\phi$, the angular separation $\Delta R = \sqrt{\Delta\eta^2 + \Delta\phi^2}$ and the total transverse momentum $p_{T,\mathrm{tot}} = p_{T1} + p_{T2}$. Events with missing values are removed globally.

The label is therefore an analytic function of a subset of the inputs. This is deliberate, because it gives a well-defined and physically meaningful target, but it also means that a large part of the achievable performance follows from the definition itself. We return to this point in Sec.~\ref{sec:limits}, since it bounds how much any of the results can be read as evidence of nontrivial pattern discovery.

All models use the same stratified split of the full dataset, with 80 percent for training and 20 percent for testing, and 10 percent of the training portion held out for validation. The quantum support vector machine uses 10{,}000 events because of the quadratic cost of kernel construction. All runs share the random seeds 21, 42 and 85 and then we take mean $\pm$ std of all metrics. Standardization parameters are fit on the training set and applied to validation and test. Scores are converted to decisions at a fixed operating threshold of 0.5. Neural models are trained for 25 epochs and their learning curves are reported. For the support-vector methods the convex problem is solved directly, with a precomputed quantum kernel feeding a classical solver in the quantum case, and convergence is summarized through hinge-loss diagnostics on the training, validation and test partitions. Performance is reported through accuracy, ROC-AUC, F1-score, precision and recall, so that ranking quality and the balance between purity and efficiency are both visible.

\section{Methods}\label{sec:methods}

\subsection{Classical models}\label{sec:classical}

\begin{figure}[h]
\centering
\includegraphics[width=0.6\textwidth]{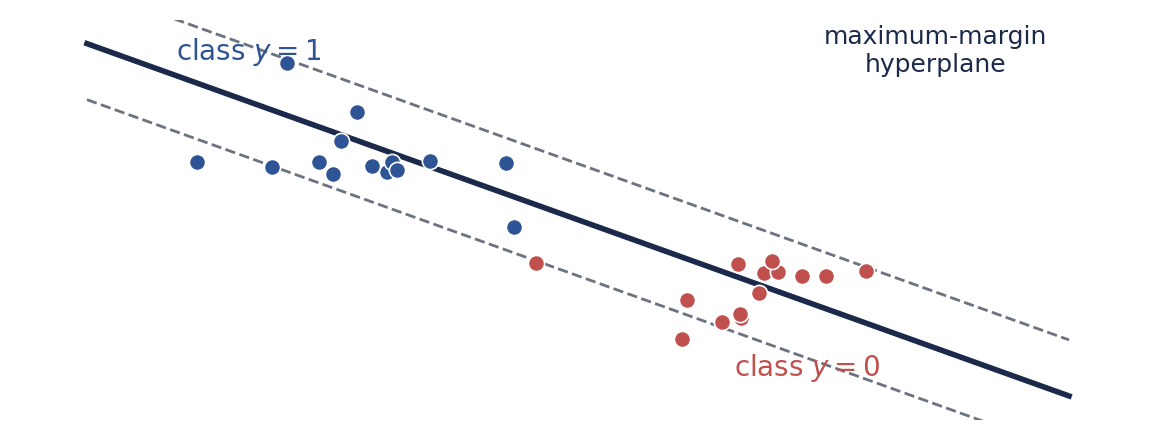}
\caption{Schematic of the support vector machine: a maximum-margin hyperplane separating the two classes, with the margin fixed by the support vectors.}\label{fig:svm}
\end{figure}

The support vector machine is a maximum-margin classifier. Its decision comes from a hyperplane that maximizes the separation between the two classes, expressed through a kernel over the support vectors (Fig.~\ref{fig:svm}),
\begin{equation}
f(x) = \sum_{i=1}^{N} \alpha_i y_i K(x_i,x) + b, \qquad \hat y = \mathrm{sign}\big(f(x)\big),
\end{equation}
where $K$ is the kernel, the weights $\alpha_i$ are learned and most vanish, $y_i$ are the class labels and $b$ is a bias \cite{cortes_svm}.

A class boundary that may not be linear was modeled with a classical Support Vector Machine (SVM) classifier implemented by using \texttt{sklearn.svm.SVC} and a radial basis function (RBF) kernel, which allows data to be implicitly mapped to a higher dimensional feature space. Model training was done using a single call to \texttt{fit} (epoch-wise training was not used), and the optimal separating hyperplane was found in the RBF-induced feature space using the LIBSVM Sequential Minimal Optimization (SMO)-type solver. The final model is composed of the learned support vectors (a subset of the training data defining the margin), the dual coefficients (Lagrange multipliers) and the bias/intercept term. The internal calibration step (Platt scaling) applied to the SVM decision scores (with the corresponding calibration parameters learned during training, by using the argument \texttt{probability=True} was used to calculate the posterior class probabilities. The threshold for inference was set to $0.5$, and if $p(y=1 \mid x) \ge 0.50$ and class $0$ otherwise. The hyperparameters of SVM were optimized as follows: RBF kernel, with $C=100$ and $\gamma=0.1$; a fixed \texttt{random\_state} (or seed for random number generator) was selected to minimize the stochastic aspects of probability calibration.

\begin{figure}[h]
\centering
\includegraphics[width=0.75\textwidth]{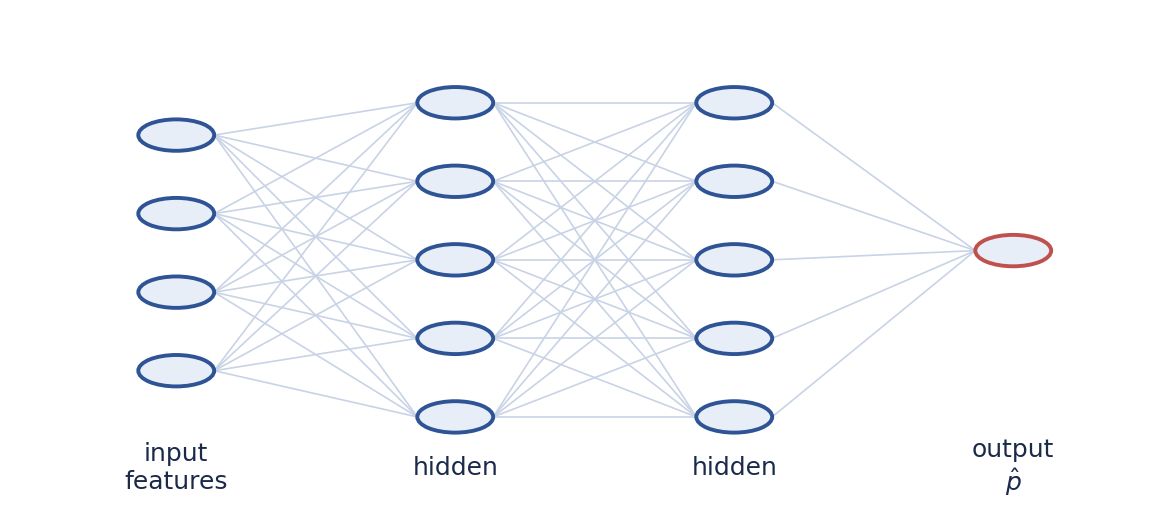}
\caption{Fully connected network for binary classification. Each hidden unit applies a nonlinear activation and the output unit returns the trigger probability.}\label{fig:ann}
\end{figure}

The artificial neural network is a stack of fully connected layers with nonlinear activations that learns a smooth mapping from features to a score (Fig.~\ref{fig:ann}),
\begin{equation}
a^{l} = \sigma\big(W^{l} a^{l-1} + b^{l}\big), \quad a^{0} = x, \quad \hat p = \sigma\big(w^{\top} a^{L} + b\big),
\end{equation}
where $W^{l}$ and $b^{l}$ are learnable, $\sigma$ is a nonlinearity such as the rectified linear unit in the hidden layers and a sigmoid at the output, and $\hat p$ estimates the probability that the trigger label equals one \cite{nair_relu}.

For the binary classification of the target variable \textit{trigger}, a feed-forward artificial neural network (ANN) was implemented in \texttt{sklearn.neural\_network.MLPClassifier}. The model was built as a fully connected multilayer perceptron (MLP) of a succession of dense hidden layers with gradually decaying dimensions to learn the more complex feature representations. Certainly, the architecture of the network is inspired from the following: Input $\rightarrow$ 256 $\rightarrow$ 128 $\rightarrow$ 64 $\rightarrow$ 32 $\rightarrow$ Output. The latter is a layer of binary classification that gives the class prediction.

\begin{figure}[h]
\centering
\includegraphics[width=0.8\textwidth]{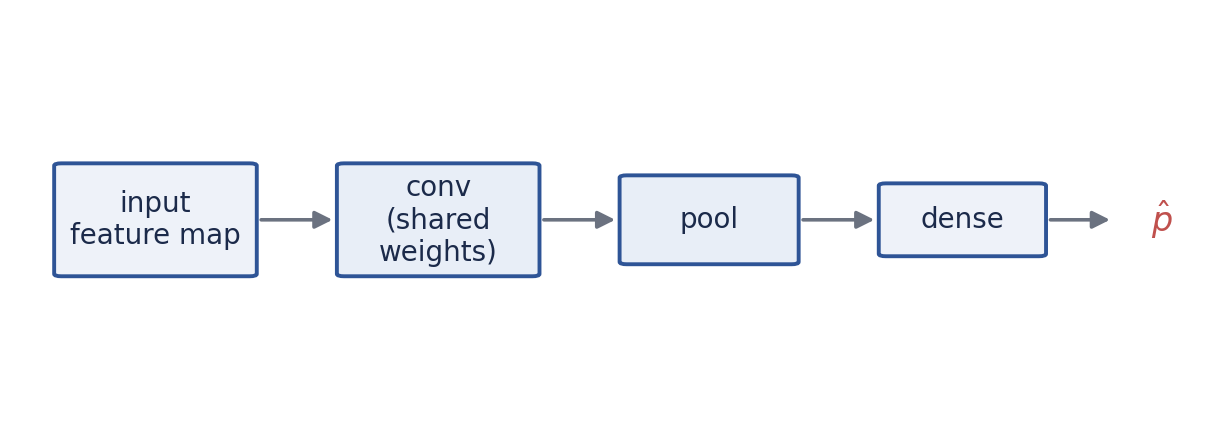}
\caption{Convolutional pipeline. Shared-weight filters and pooling extract local patterns before a dense head produces the score.}\label{fig:cnn}
\end{figure}

The convolutional network learns filters with shared weights that exploit local structure (Fig.~\ref{fig:cnn}),
\begin{equation}
F^{l}_{i,j} = \sigma\Big(\sum_{m,n} K_{m,n}\, F^{l-1}_{i+m,j+n} + b\Big),
\end{equation}
where $K_{m,n}$ is a trainable filter applied at every position, $F^{l-1}$ is the input map and $\sigma$ is the nonlinearity \cite{oshea_cnn}.

A one dimensional convolutional neural network (1D-CNN) framework was used to implement a  binary classification system with PyTorch. It is a tabular feature vector of length $L = n_{\text{features}}$, which is reshaped into a single channel 1D sequence ($x \in \mathbb{R}^{L} \rightarrow x' \in \mathbb{R}^{1 \times L}$) for convolutional processing. The backbone comprises three convolutional blocks: (i) $\text{Conv1d}(1 \rightarrow 128,\; k=3,\; \text{padding}=1)$ followed by $\text{BatchNorm1d}(128)$ and the SiLU activation; (ii) $\text{Conv1d}(128 \rightarrow 256,\; k=3,\; \text{padding}=1)$ followed by $\text{BatchNorm1d}(256)$ and SiLU; and (iii) $\text{Conv1d}(256 \rightarrow 512,\; k=3,\; \text{padding}=1)$ followed by $\text{BatchNorm1d}(512)$ and SiLU. It is an aggregate of the features across the entire image, averaged by the function using $\text{AdaptiveAvgPool1d}(1)$, implemented as a 1x1 pooled value per channel. The classifier head comprises a multi-layer perceptron with one logit, which has layers of $\text{Linear}(512 \rightarrow 512)$ + SiLU + Dropout(0.4), $\text{Linear}(512 \rightarrow 256)$ + SiLU + Dropout(0.3), and a final $\text{Linear}(256 \rightarrow 1)$. The output of the network is in the form of a logit $z$ which is passed to a sigmoid function for the output to be a posterior probability $p=\sigma(z)$ and a threshold $0.5$ used to make the decision on what the network is predicting (class $1$, or class $0$). The size of the batch was set as 1024 and optimizer \texttt{AdamW} was used for the training. All parameters of all convolutional kernels, biases, all the weights and biases of the fully connected layers were trainable, as well as all the parameters of the backbone (including the parameters of the batch normalization layers (scale and shift), in the case of batch-norm, the running stats were updated during training).

\begin{figure}[h]
\centering
\includegraphics[width=0.78\textwidth]{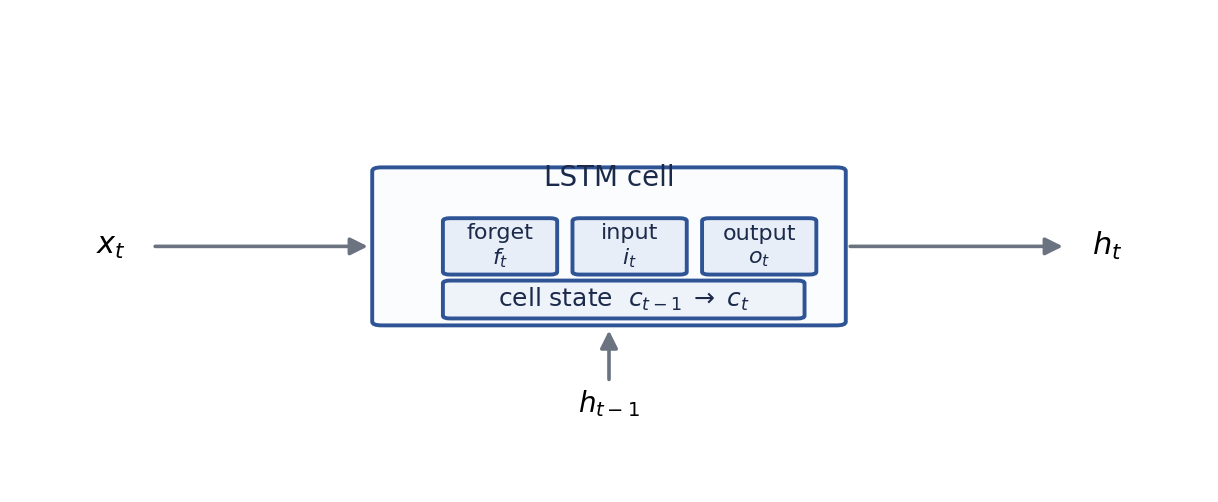}
\caption{Long short-term memory cell, with the input, forget and output gates controlling the flow into and out of the memory cell.}\label{fig:lstm}
\end{figure}

The long short-term memory network is a gated recurrent model that controls what to remember and what to forget (Fig.~\ref{fig:lstm}). With $z_t = [h_{t-1}, x_t]$,
\begin{align}
i_t &= \sigma(W_i z_t + b_i), \quad f_t = \sigma(W_f z_t + b_f), \quad o_t = \sigma(W_o z_t + b_o), \nonumber \\
\tilde c_t &= \tanh(W_c z_t + b_c), \quad c_t = f_t \odot c_{t-1} + i_t \odot \tilde c_t, \quad h_t = o_t \odot \tanh(c_t),
\end{align}
where $i_t$, $f_t$ and $o_t$ are the input, forget and output gates, $c_t$ is the memory cell, $h_t$ is the hidden state and $\odot$ denotes element-wise multiplication \cite{hochreiter_lstm}.

The binary classification was done using a stacked Long Short-Term Memory (LSTM) network, which was implemented in PyTorch. The input vector $x$ of the original tabular data with $n_{\text{feature}}$ variables was reshaped and fed as an input sequence of length one ($x \in \mathbb{R}^{n_{\text{features}}} \rightarrow x' \in \mathbb{R}^{1 \times n_{\text{features}}}$) to the recurrent input interface. LSTM network (LSTMNet) consists of two LSTM layers that are sequentially connected, and is applied with \texttt{batch\_first=True}. The size of input features in the first LSTM layer is $n_{\text{feature}}$, the size of hidden features is 128, and the regularization size of the sequence output is Dropout(0.3). The 128-dimensional sequence representation is fed to the second LSTM layer which outputs a 256-dimensional hidden representation that is again passed to another Dropout(0.3). The MLP head used for classification applied to the last time step representation $z[:, -1, :] \in \mathbb{R}^{256}$, comprising $\text{Linear}(256 \rightarrow 128)$ + SiLU, $\text{Linear}(128 \rightarrow 64)$ + SiLU, and a final $\text{Linear}(64 \rightarrow 1)$ that produces a single logit. The logit $z$ is transformed to a posterior probability using the sigmoid function $p = \sigma(z)$ and predictions are obtained by applying a threshold value of 0.5 (class $1$ if $p \ge 0.50$, class 0). All the weights and biases of both LSTM layers (input-to-hidden and hidden-to-hidden matrices and the biases on the gates in the standard LSTM formulation) and of the MLP head were trainable, while dropout layers added no trainable parameters. \texttt{AdamW} was used for optimization with batch size 1024 for the model.

\subsection{Quantum models}\label{sec:quantum}

Every quantum model works under the same budget: eight qubits, a principal-component compression of the inputs to sixteen features, and state-vector simulation. The compression maps the standardized features onto sixteen components, which are then encoded into the eight-qubit register.

\begin{figure}[h]
\centering
\includegraphics[width=0.82\textwidth]{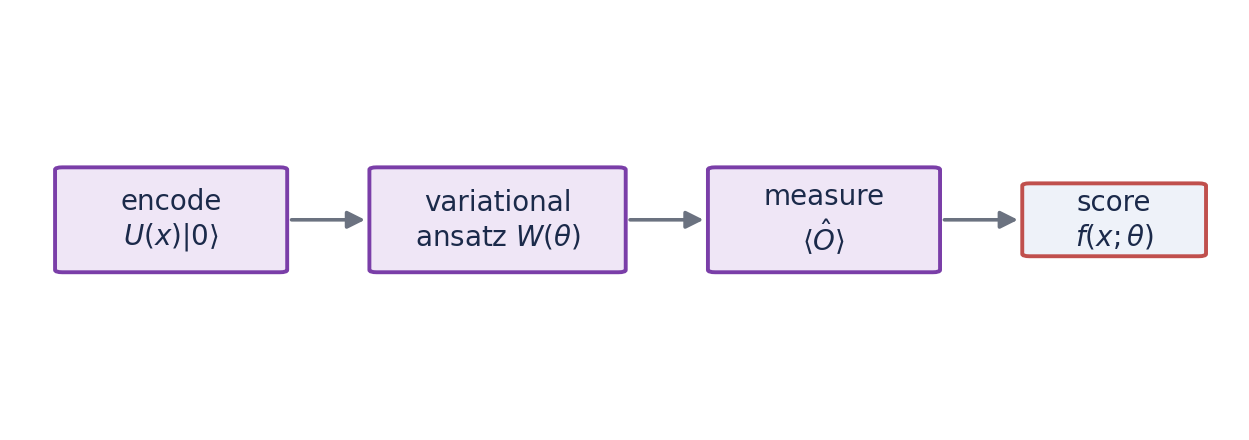}
\caption{Quantum neural network: data encoding, a variational ansatz and a measurement that returns the score.}\label{fig:qnn}
\end{figure}

The quantum neural network is a variational circuit that encodes the classical features into a quantum state, applies a trainable ansatz and reads out an expectation value (Fig.~\ref{fig:qnn}),
\begin{equation}
f(x;\theta) = \langle 0 | U^{\dagger}(x)\, W^{\dagger}(\theta)\, \hat O\, W(\theta)\, U(x) | 0 \rangle,
\end{equation}
where $U(x)$ encodes the data, $W(\theta)$ is the trainable ansatz and $\hat O$ is the measured observable whose expectation is mapped to a classification score \cite{farhi_qnn}.

The PennyLane framework was used to implement a hybrid quantum-classical binary classifier with the \texttt{lightning.qubit} simulator backend that is statevector based. The quantum circuit (QNode) was integrated with PyTorch, and was trained end-to-end using efficient gradient computation method  adjoint differentiation. Properly PCA reducing input features to $ N_{\text{PCA}}$ PCA components and then transforming them with $ \arctan(\cdot)$, to quantum rotation angles. All the PCA components were re-uploaded into the data using a re-uploading strategy where the number of re-uploads was given as $N_{\text{UPLOADS}} = N_{\text{PCA}}/N_{\text{QUBITS}}$ (with $N_{\text{QUBITS}}=8$); for the embedding function used, \texttt{AngleEmbedding} with Y-rotations. The trainable part of the variational layer was a \texttt{StronglyEntanglingLayers} with a depth of $ Q_{\text{LAYERS}}=1$ layers per upload, for a total $ N_{\text{UPLOADS}} \times 1 \times 8 \times 3$ trainable parameters at the quantum layer. The result from the readout was a list of expectation values of single-qubit measurements $\langle Z_i\rangle$ and $\langle X_i\rangle$ for $i=0,\ldots,7$, and a list of correlators of nearest neighbors of the type $\langle Z_i Z_{i+1}\rangle$ for $i=0,\ldots,6$,, giving a quantum feature dimension of $ Q_{\text{OUT}} = 2N_{\text{QUBITS}} + (N_{\text{QUBITS}}-1) = 23$. A classical MLP head consisting of the following layers was used: $\text{Linear}(N_{\text{PCA}}+Q_{\text{OUT}} \rightarrow 128)$, $\text{Linear}(128 \rightarrow 64)$, and $\text{Linear}(64 \rightarrow 1)$. The classical MLP head was trained with the classical PCA feature vector and the concatenated quantum features with dropout regularization, but dropout was not added to the model as a layer to train. The \texttt{BCEWithLogitsLoss} was used to train the model, using \texttt{AdamW} with a learning rate of ($3\times10^{-3}$) for the quantum layer and a learning rate of ($2\times10^{-3}$) for the classical head. The mini-batch size was set to 256 for training and 512 for validation/testing and shuffling the data was performed using a seeded generator to guarantee the reproducibility.

\begin{figure}[h]
\centering
\includegraphics[width=0.82\textwidth]{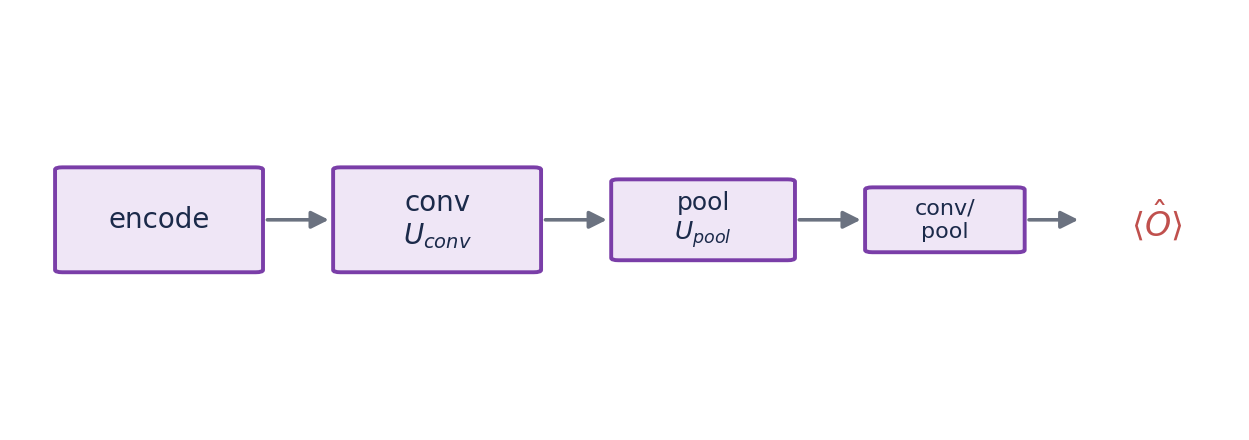}
\caption{Quantum convolutional network, with alternating convolution and pooling unitaries followed by a measurement.}\label{fig:qcnn}
\end{figure}

The quantum convolutional network stacks convolution-like unitaries with pooling blocks that reduce the number of active qubits, in analogy with a classical convolutional network (Fig.~\ref{fig:qcnn}),
\begin{equation}
|\psi_{\mathrm{out}}\rangle = \prod_{s=1}^{S} U^{s}_{\mathrm{pool}}(\theta)\, U^{s}_{\mathrm{conv}}(\theta)\, U(\theta)\, |0\rangle, \qquad f(x;\theta) = \langle \psi_{\mathrm{out}} | \hat O | \psi_{\mathrm{out}} \rangle,
\end{equation}
where $U_{\mathrm{conv}}$ acts locally on qubits and $U_{\mathrm{pool}}$ reduces the register before the observable is measured \cite{chen_qcnn}.

A hybrid quantum–classical QCNN architecture, combined with PennyLane, was used to solve the binary classification problem, using the \texttt{lightning.qubit} statevector simulator. The adjoint differentiation method was used to implement the quantum node (QNode) in PyTorch and train it. The classical features are coded into the quantum rotations angles by using the function $ \arctan(\cdot)$ and are embedded using a data re-uploading strategy, where the number of data re-uploads is given by $N_{\text{UPLOADS}} = N_{\text{PCA}}/N_{\text{QUBITS}}$ (with $N_{\text{QUBITS}}=8$). The 8 angles were then attached to the 8 qubits by using Y-rotations for each one of them in the state of uploading: \texttt{AngleEmbedding}. Two-qubit “convolution” and “pooling” blocks were used to parameterize the blocks of QCNN ansatz. The convolution block (6 trainable parameters per block) applied, on a qubit pair $(a,b)$, a sequence of rotations and entangling operations: $RY(a), RY(b)$, CNOT$(a\rightarrow b)$, $RZ(a), RZ(b)$, CNOT$(b\rightarrow a)$, followed by $RY(a), RY(b)$. Per upload, one convolution round was executed using eight blocks arranged over non-overlapping pairs $(0,1),(2,3),(4,5),(6,7)$ and shifted pairs $ (1,2),(3,4),(5,6),(7,0)$. Following all the uploads, a fixed pooling stage (2 trainable parameters for each pooling block) was performed using 4 pooling blocks with topology $0\leftarrow1$, $2\leftarrow3$, $4\leftarrow5$, and $6\leftarrow7$, followed by a single-qubit rotation layer (3 parameters for each kept qubit) using $ RX/RY/RZ$ on the kept qubits \{0,2,4,6\}. Quantum features were obtained by measuring expectation values on the kept qubits, consisting of $\langle Z\rangle$ and $\langle X\rangle$ on qubits 0, 2, 4, 6 (8 values total) and nearest-neighbor correlators $\langle Z_0Z_2\rangle$, $\langle Z_2Z_4\rangle$, and $\langle Z_4Z_6\rangle$ (3 values), yielding $Q_{\text{CNN\_OUT}}=11$ quantum outputs. These quantum features were concatenated with the PCA feature vector and fed to a classical MLP head, $\text{Linear}(N_{\text{PCA}}+11 \rightarrow 128)$ + SiLU + Dropout(0.25), $\text{Linear}(128 \rightarrow 64)$ + SiLU + Dropout(0.15), and $\text{Linear}(64 \rightarrow 1)$ to produce a single logit. All of the model parameters (quantum circuit parameters, head weights and biases) were end-to-end trained by minimizing the \texttt{BCEWithLogitsLoss} with the optimizer \texttt{AdamW} (with two parameter groups, with the same learning rate as in above Quantum models).

\begin{figure}[h]
\centering
\includegraphics[width=0.82\textwidth]{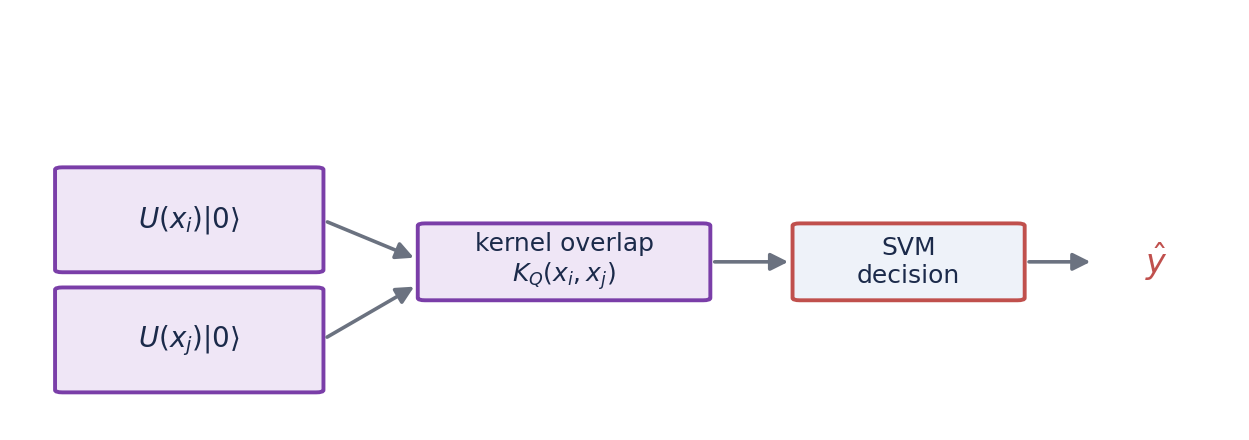}
\caption{Quantum support vector machine: a state-overlap kernel feeds a classical margin classifier.}\label{fig:qsvm}
\end{figure}

The quantum support vector machine replaces the classical kernel with an overlap between encoded states (Fig.~\ref{fig:qsvm}),
\begin{equation}
K_Q(x_i,x_j) = \big| \langle 0 | U^{\dagger}(x_i) U(x_j) | 0 \rangle \big|^2, \qquad f(x) = \sum_{i=1}^{N} \alpha_i y_i K_Q(x_i,x) + b,
\end{equation}
with $\hat y = \mathrm{sign}(f(x))$. The parameters $\alpha_i$, $y_i$ and $b$ keep their classical meaning, now defined in a quantum feature space \cite{schuld_hilbert,wu_qsvm}.

A kernel Support Vector Machine (QSVM) was implemented using PennyLane and the statevector simulator from the \texttt{lightning.qubit} package. This quantum circuit was not optimized with gradient descent as in variational quantum classifiers, but was just assumed to be a fixed quantum feature map that specifies the kernel depending on the data. Features were mapped to bounded rotation angles (roughly in the rage $ (-\pi,\pi) $, and embedded using data re-uploading scheme of $N_{\text{UPLOADS}} = N_{\text{PCA}}/N_{\text{QUBITS}}$ for $N_{\text{QUBITS}}=8$. Each upload, it used the \texttt{AngleEmbedding} with Y-rotations on all qubits, then one iteration of \texttt{StronglyEntanglingLayers} with parameters sampled at random from $ \mathcal{N}(0,0.35) $ but not optimized (fixed). The circuit output was the full statevector $ |\psi(x)\rangle \in \mathbb{C}^{2^{8}} $ (256 dimensional vector space) which was used to define a quantum fidelity kernel between two inputs $x$ and $x'$ as $ K(x,x') = |\langle \psi(x)\,|\,\psi(x')\rangle|^{2} $. In practice, the kernel Gram matrix was computed from the simulated statevectors (e.g., using ($ G = A A^{\dagger} $) with rows of $A$ being statevectors of the data and fed to a classical SVM using a precomputed kernel. The SVM then trained the standard dual representation, the support vectors (support indices), the dual coefficients as well as the intercept term. This has been done with the use of \texttt{probability=True} which required an extra calibration step to generate \texttt{predict\_proba} outputs, and classification was done using the threshold of 0.5. The precomputed kernel matrix and the regularization parameter $C=1.0$ were used to train the QSVM.

\begin{figure}[h]
\centering
\includegraphics[width=0.82\textwidth]{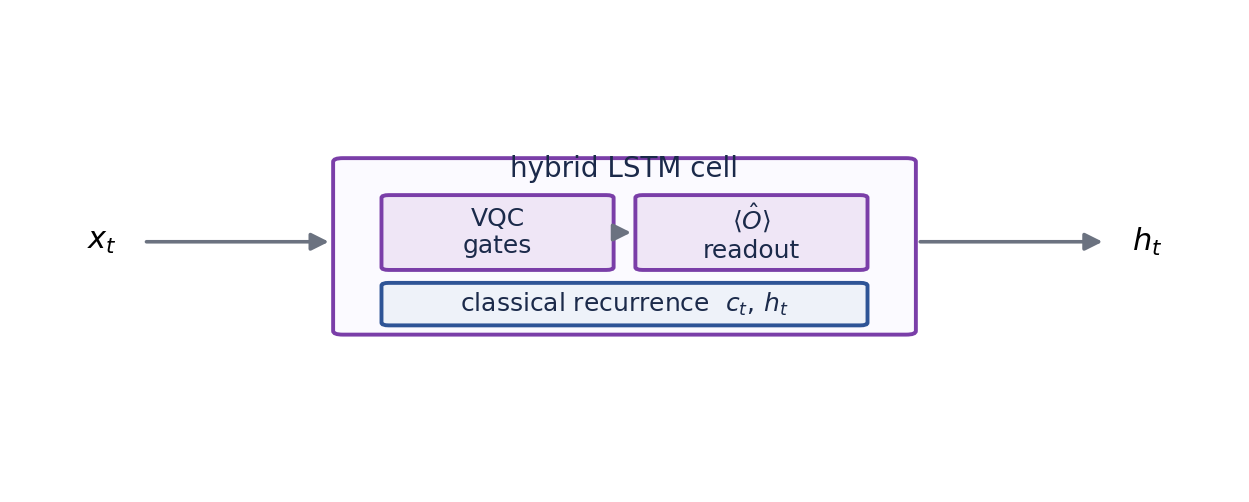}
\caption{Hybrid long short-term memory cell, in which a variational circuit produces the gate pre-activations and the recurrence remains classical.}\label{fig:qlstm}
\end{figure}

The quantum long short-term memory network computes the gate pre-activations with a variational circuit and keeps the recurrence classical (Fig.~\ref{fig:qlstm}). With $z_t = [h_{t-1}, x_t]$,
\begin{align}
\tilde g_t &= \langle \hat O \rangle_{\mathrm{VQC}}\big(z_t; \theta_g\big), \nonumber \\
c_t &= f_t \odot c_{t-1} + i_t \odot \tilde g_t, \quad h_t = o_t \odot \tanh(c_t),
\end{align}
where the pre-activations $\tilde g_t$ are read from the circuit and the classical memory update is applied to them \cite{chen_qlstm}.

Using the quantum backend, qubit simulator \texttt{lightning.qubit}, a PyTorch/PennyLane hybrid quantum long short term memory (QLSTM) model was implemented. The quantum node (QNode) was set up using torch interface  and trained end-to-end using adjoint differentiation. The classical input was then projected onto an 8-dimensional angle vector by a learnable linear matrix \texttt{Linear(input\_dim + hidden\_dim} $\rightarrow$ \texttt{8)} (bounded approximately within $[-\pi/2,\pi/2]$) and the quantum circuit input was the concatenation of the current classical input vector $x_t$ with the previous hidden state of the quantum circuit $ h_{t-1} $. The variational ansatz \texttt{StronglyEntanglingLayers} with $ Q_{\text{LAYERS}}=1 $ was then applied to the Y-rotations that were done on $ N_{\text{QUBITS}}=8 $ qubits with \texttt{AngleEmbedding}. The expectation value for the features $\langle Z_i\rangle$ and $\langle X_i\rangle$ for $i=0,\ldots,7$ and the quantum size of the layer is 16 features per time step, $ Q_{\text{OUT}}=16 $. The weights of the quantum layers \texttt{StronglyEntanglingLayers} were the only weights that were trainable, and were stored in a TorchLayer $ (Q_{\text{LAYERS}}, N_{\text{QUBITS}}, 3)=(1,8,3) $ (meaning the same quantum layer was used at each time step, these weights were thus shared across all the samples and all the time steps). Typical affine transformations in LSTM were not employed for computing the activations of the gates $ (f_t,i_t,o_t,g_t) $, but rather quantum features were. Four layers of linear networks were trained, one for each type of gate, to map a 16-dimensional (16-D) feature vector from the quantum circuit to a 16-D vector for forget, input, output and candidate gates, respectively. The model is trained with \texttt{BCEWithLogitsLoss}, and using batch sizes of 512 for training and 1024 for evaluation and optimized by \texttt{AdamW}. 

\section{Results}\label{sec:results}

\subsection{Classical models}\label{sec:res-classical}

Table~\ref{tab:classical} collects the classical baselines. The artificial neural network is the strongest model, at 93.53 percent accuracy and 0.9819 ROC-AUC, with a well-balanced precision of 0.9104 and recall of 0.9458. The convolutional and recurrent networks follow closely, both near 90 percent accuracy and with areas under the curve of 0.9629 and 0.9606. The support vector machine keeps a fairly high recall of 0.8624 but a lower accuracy of 77.38 percent and an area under the curve of 0.8186, which places it clearly behind the neural architectures.

\begin{table}[h]
\caption{Classical models on the full dataset for the trigger-like task.}\label{tab:classical}
\begin{tabular}{@{}lccccc@{}}
\toprule
Model & Accuracy & F1-score & Precision & Recall & ROC-AUC \\
\midrule
ANN  & 93.53 & 0.9278 & 0.9104 & 0.9458 & 0.9819 \\
CNN  & 89.98 & 0.8899 & 0.8612 & 0.9205 & 0.9629 \\
LSTM & 89.86 & 0.8894 & 0.8545 & 0.9273 & 0.9606 \\
SVM  & 77.38 & 0.7703 & 0.6959 & 0.8624 & 0.8186 \\
\botrule
\end{tabular}
\end{table}

\begin{figure}[h]
\centering
\includegraphics[width=0.85\textwidth]{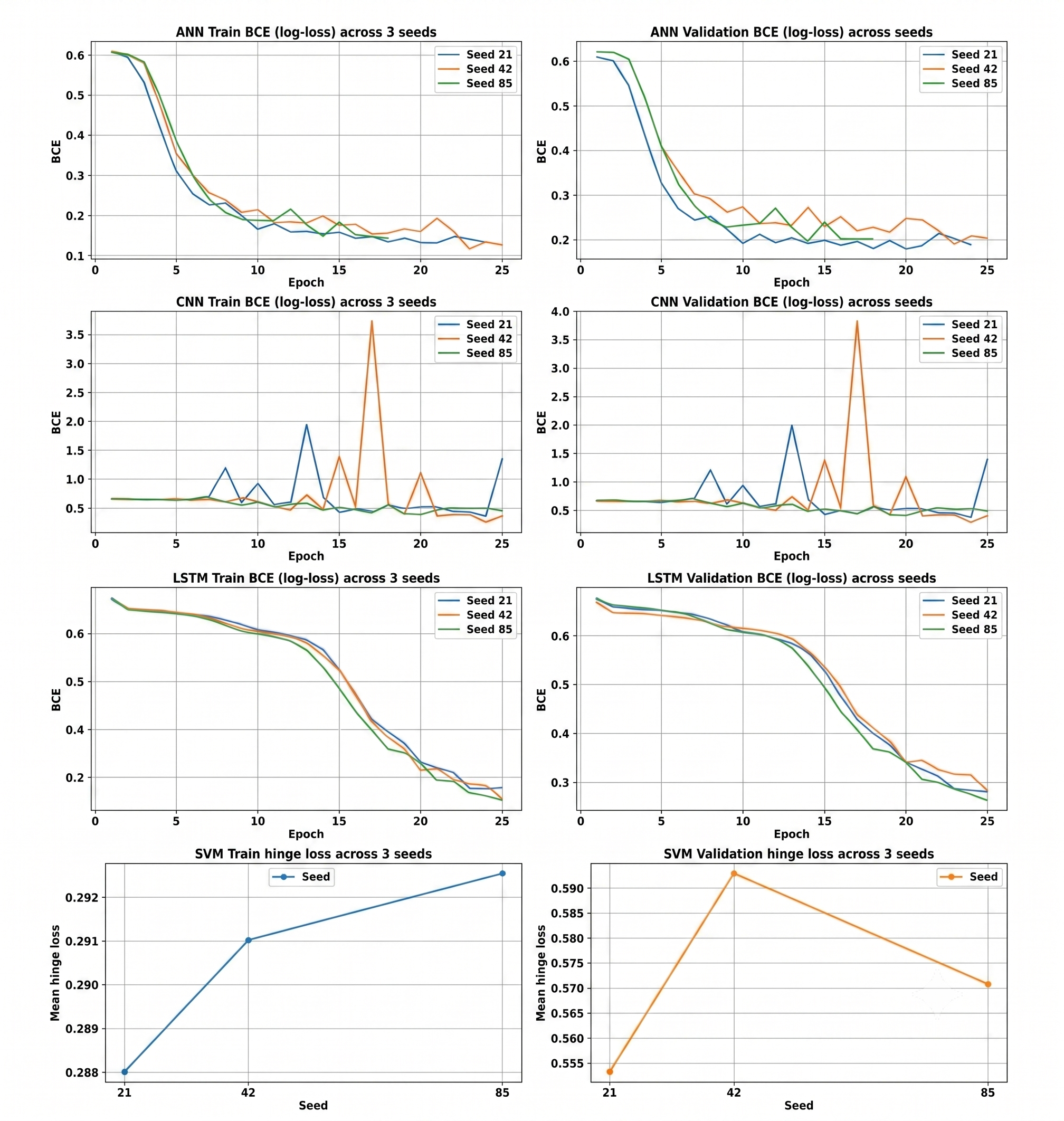}

\caption{Training and validation loss for the ANN, CNN and LSTM, with the hinge-loss diagnostic bars for the SVM.}\label{fig:loss-classical}
\end{figure}

\begin{figure}[h]
\centering
\includegraphics[width=\textwidth]{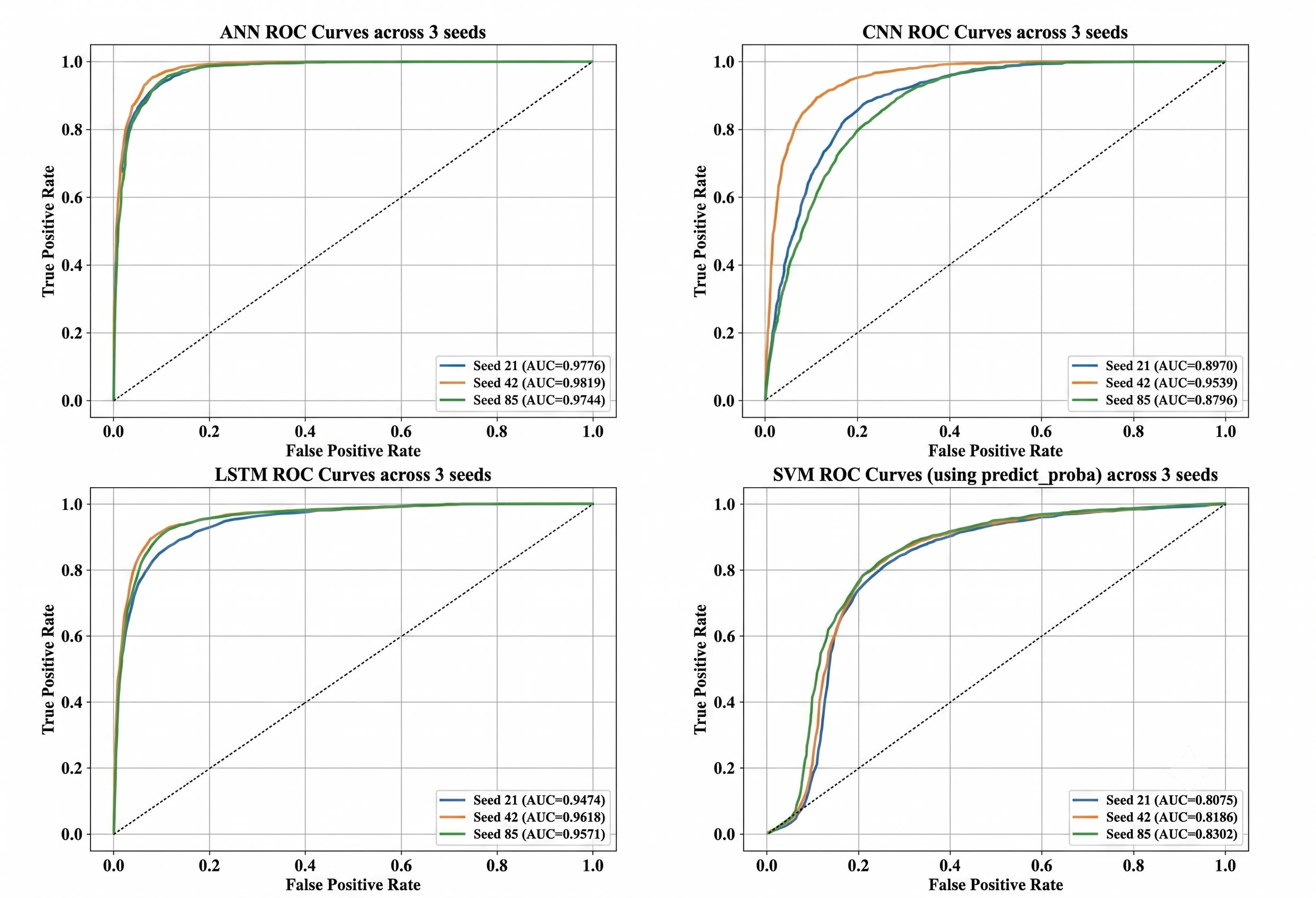}

\caption{ROC curves for the classical models across 3 seeds.}\label{fig:roc-classical}
\end{figure}

\begin{figure}[h]
\centering
\includegraphics[width=\textwidth]{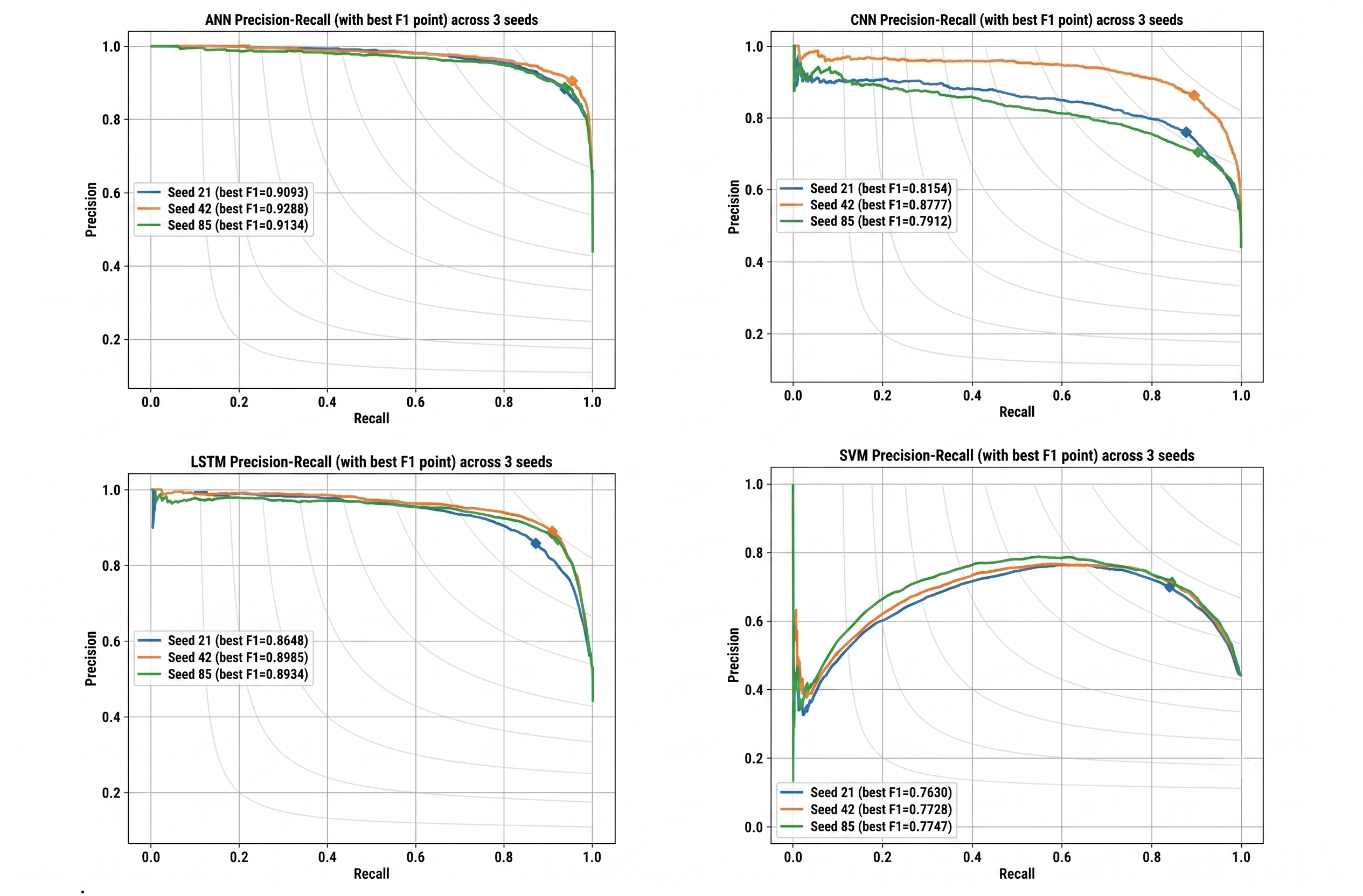}

\caption{F1-score comparison for the classical models.}\label{fig:f1-classical}
\end{figure}

\begin{figure}[h]
\centering
\includegraphics[width=0.85\textwidth]{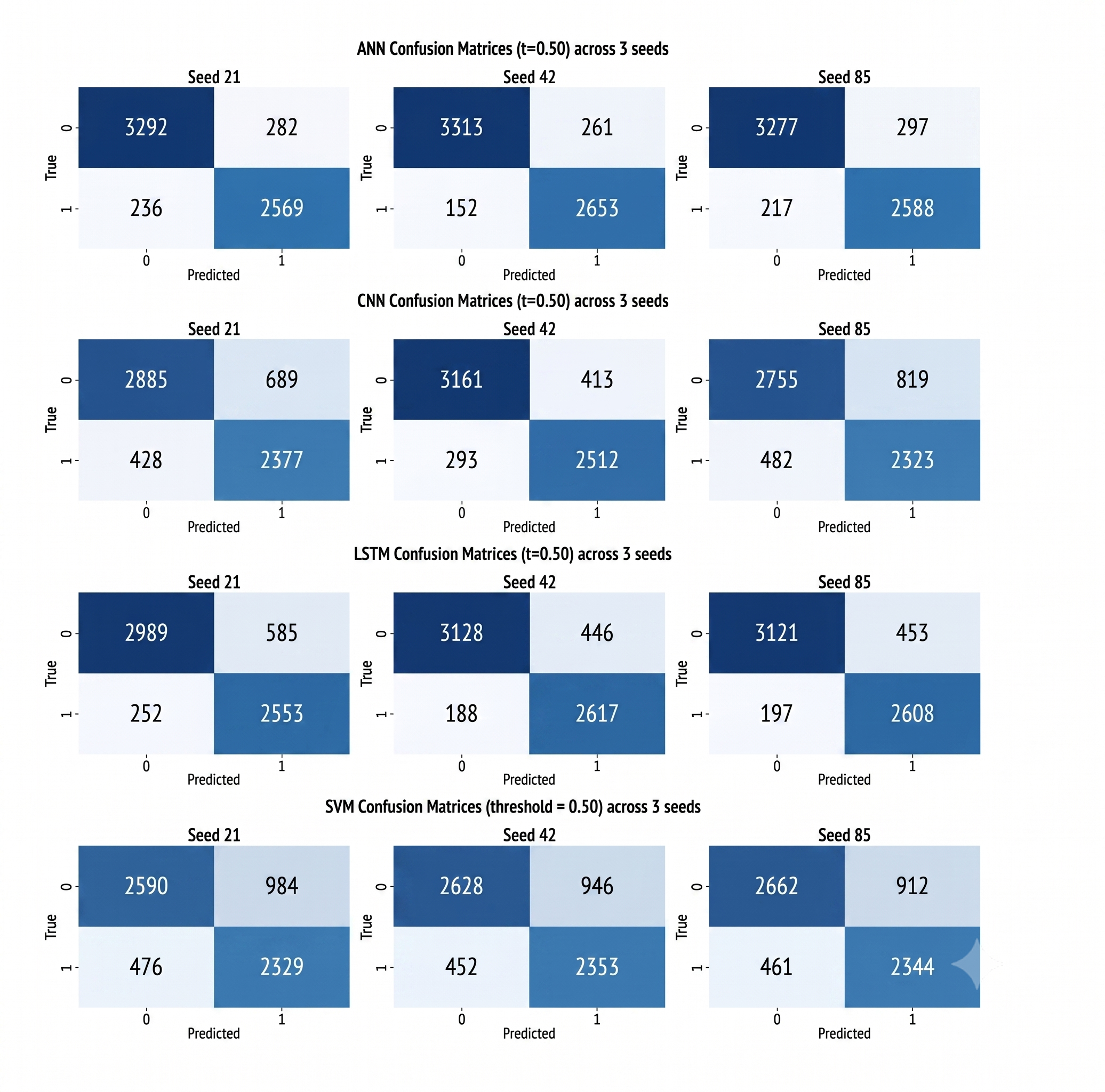}

\caption{Confusion matrices for the classical models at the fixed threshold of 0.5.}\label{fig:cm-classical}
\end{figure}

The learning curves for the neural models, together with the hinge-loss diagnostic for the support vector machine, are shown in Fig.~\ref{fig:loss-classical}. The smooth validation behavior of the artificial neural network is consistent with its accuracy and with stable convergence. The ROC curves and the F1-scores in Figs.~\ref{fig:roc-classical} and~\ref{fig:f1-classical} tell the same story: the artificial neural network leads, the convolutional and recurrent networks sit just behind at an area near 0.96, and the support vector machine trails. The confusion matrices in Fig.~\ref{fig:cm-classical} confirm that the artificial neural network combines the highest precision and recall at the chosen operating point, while the support vector machine accepts more false positives.

\subsection{Quantum models}\label{sec:res-quantum}

Table~\ref{tab:quantum} collects the quantum models. The quantum convolutional network is the strongest, at 90.89 percent accuracy and 0.9731 ROC-AUC, with the quantum neural network close behind at 90.58 percent and 0.9706. Both reach recall above 0.95, which corresponds to high signal efficiency at the chosen operating point. The other two are markedly weaker: the quantum long short-term memory network reaches 74.07 percent accuracy and 0.8325 ROC-AUC, and the quantum support vector machine reaches 67.86 percent and 0.7340. Within this budget, the trainable hybrid embeddings outperform both the fixed quantum kernel and the recurrent quantum construction.

\begin{table}[h]
\caption{Quantum models on the full dataset for the trigger-like task, under eight qubits and a principal-component compression to sixteen features.}\label{tab:quantum}
\begin{tabular}{@{}lcccccc@{}}
\toprule
Model & Qubits & PCA & Accuracy & F1-score & Precision & Recall \\
\midrule
QNN   & 8 & 16 & 90.58 & 0.8990 & 0.8505 & 0.9533 \\
QCNN  & 8 & 16 & 90.89 & 0.9022 & 0.8546 & 0.9554 \\
QLSTM & 8 & 16 & 74.07 & 0.7078 & 0.7016 & 0.7141 \\
QSVM  & 8 & 16 & 67.86 & 0.5952 & 0.6671 & 0.5373 \\
\botrule
\end{tabular}
\footnotetext{ROC-AUC values are 0.9706 (QNN), 0.9731 (QCNN), 0.8325 (QLSTM) and 0.7340 (QSVM).}
\end{table}

\begin{figure}[h]
\centering
\includegraphics[width=\textwidth]{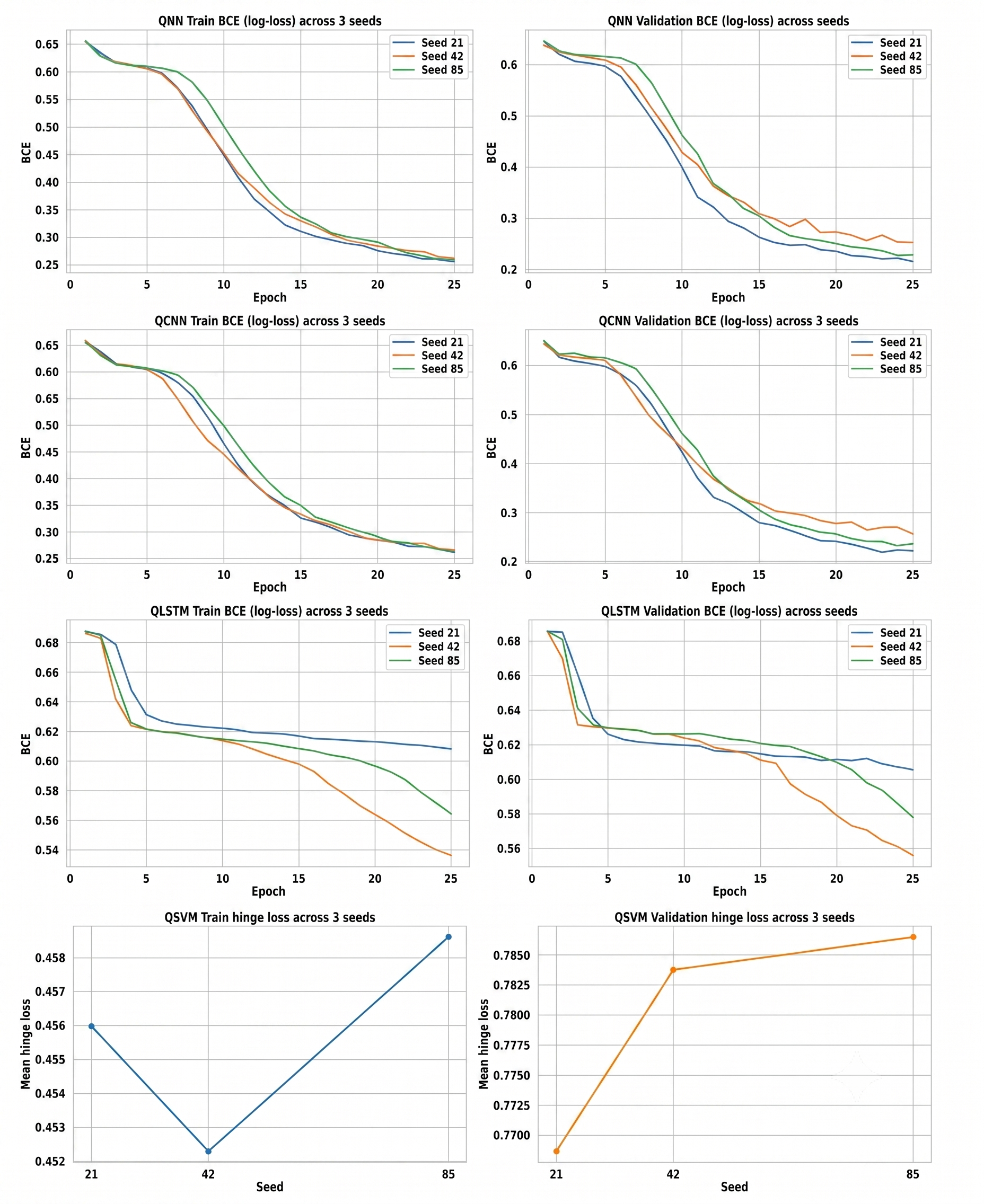}

\caption{Training and validation loss for the QNN, QCNN and QLSTM, with the hinge-loss diagnostic bars for the QSVM.}\label{fig:loss-quantum}
\end{figure}

\begin{figure}[h]
\centering
\includegraphics[width=\textwidth]{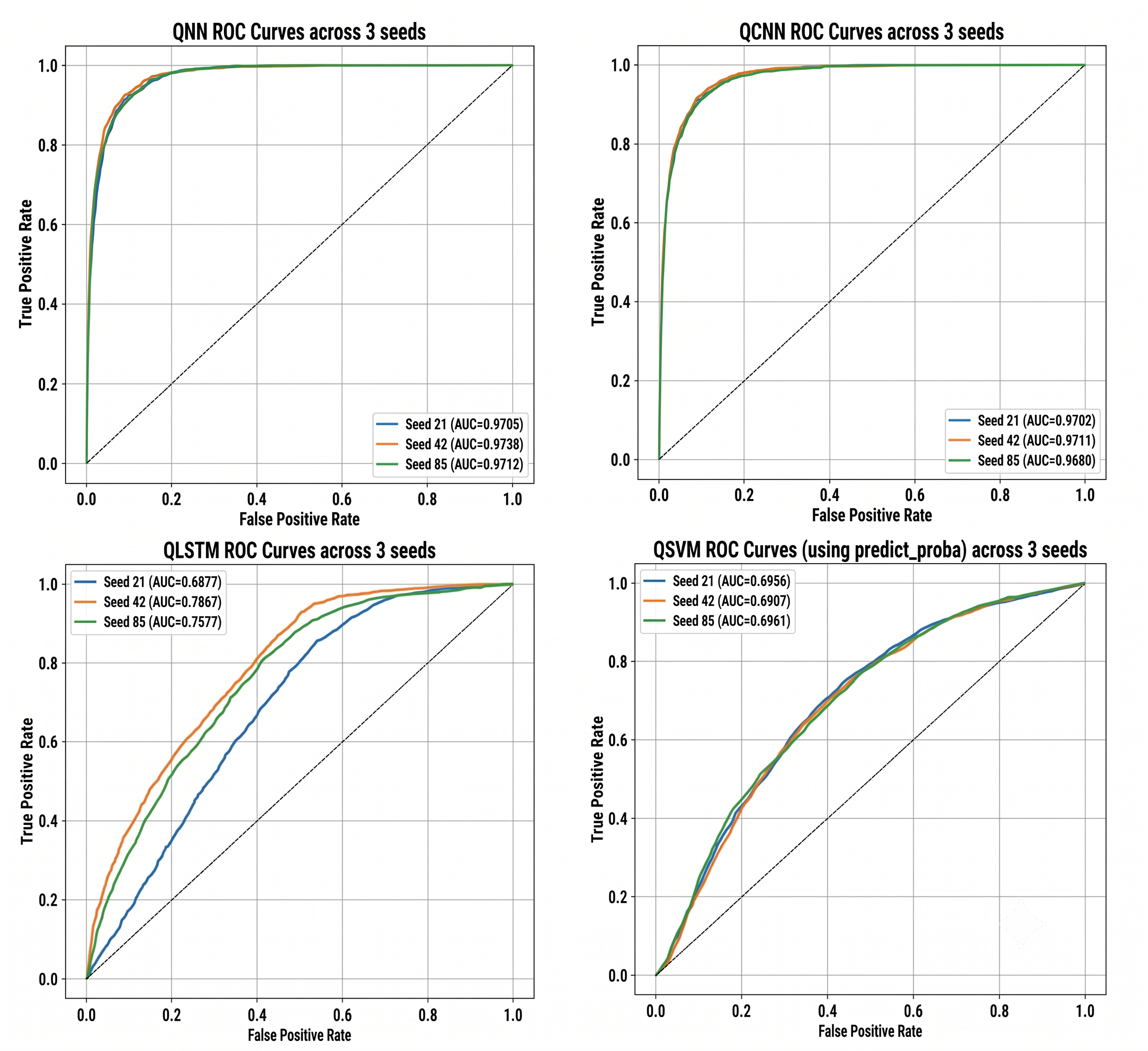}

\caption{ROC curves for the quantum models.}\label{fig:roc-quantum}
\end{figure}

\begin{figure}[h]
\centering
\includegraphics[width=\textwidth]{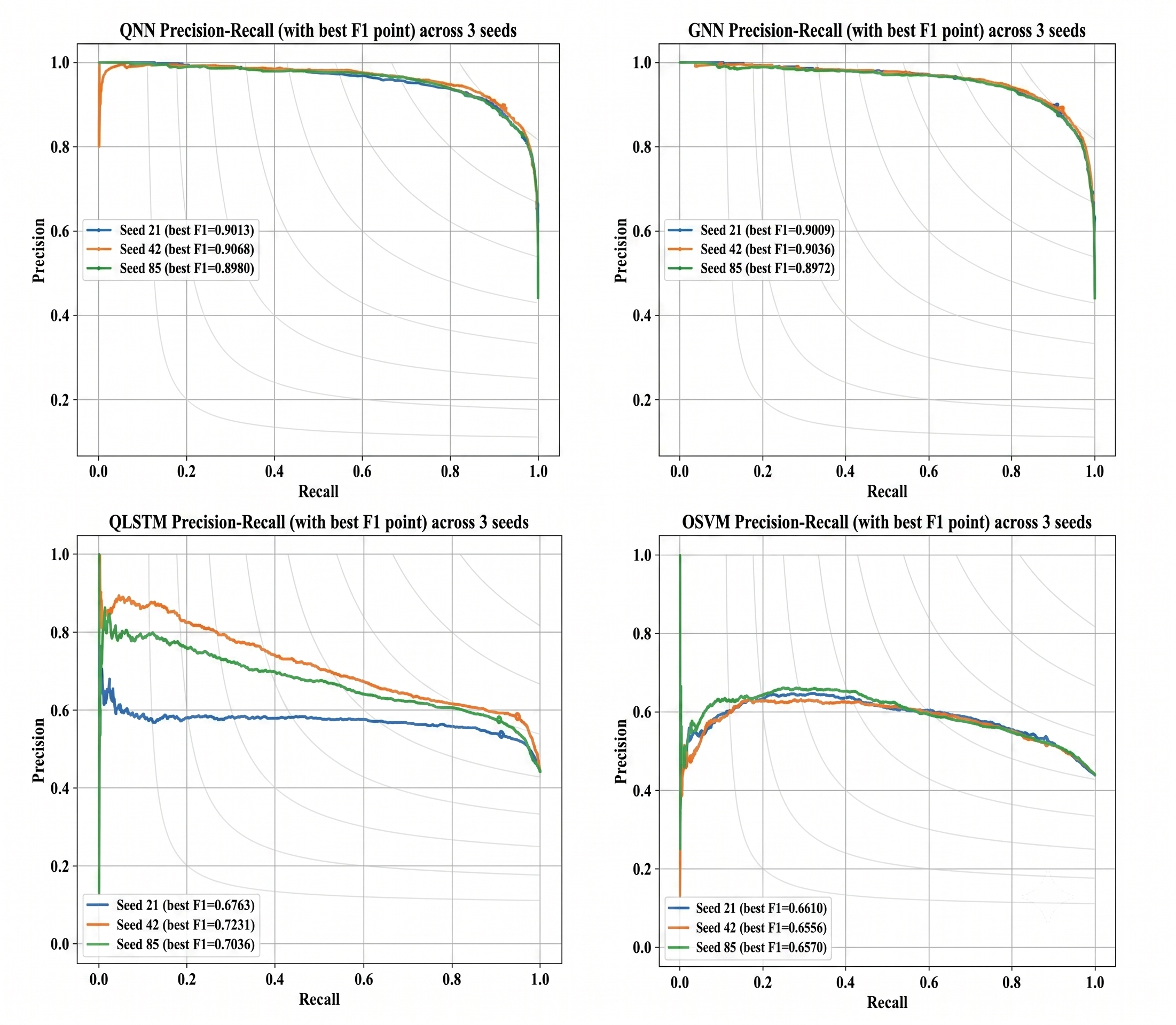}

\caption{F1-score comparison for the quantum models.}\label{fig:f1-quantum}
\end{figure}

\begin{figure}[h]
\centering
\includegraphics[width=0.85\textwidth]{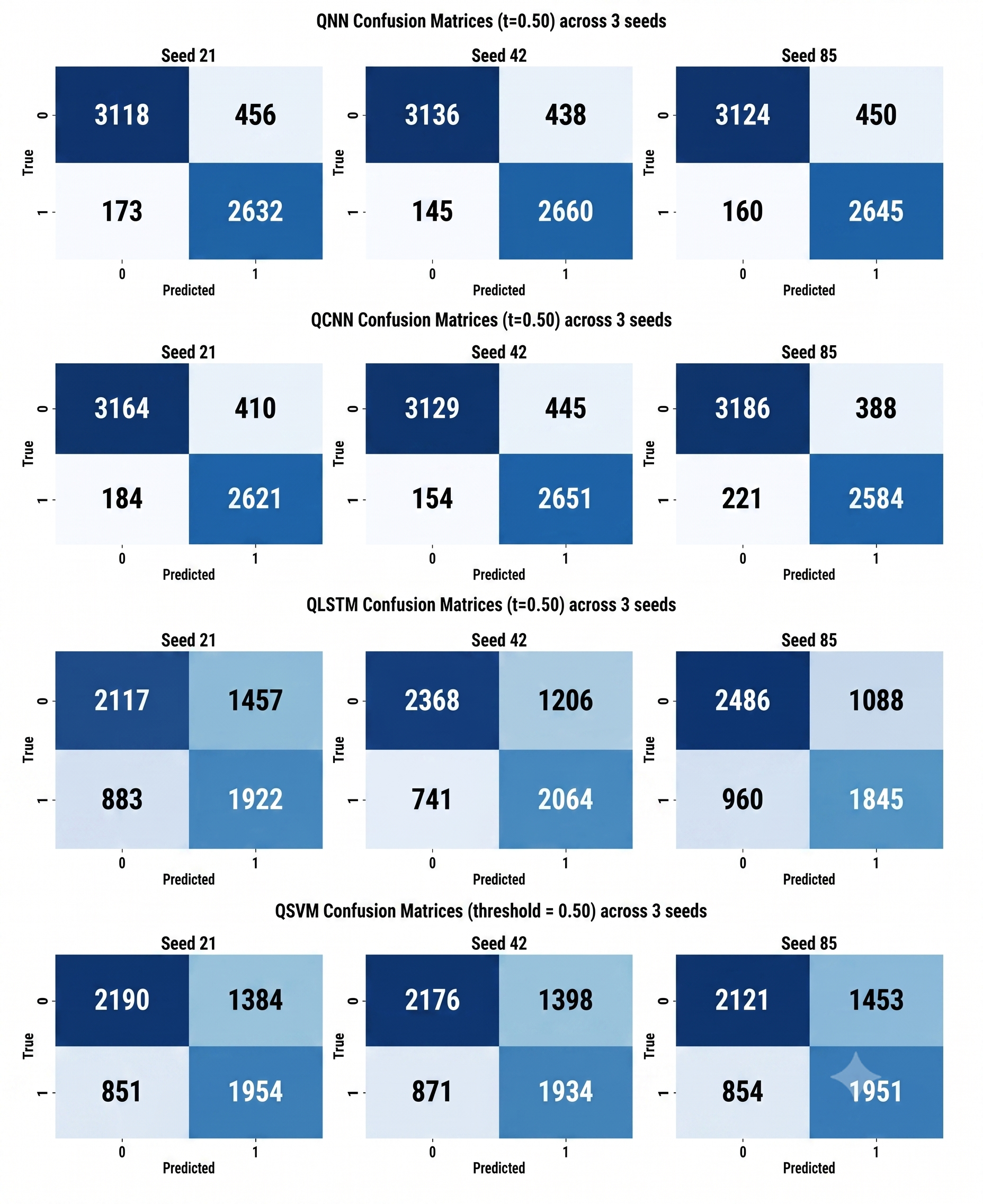}

\caption{Confusion matrices for the quantum models at the fixed threshold of 0.5.}\label{fig:cm-quantum}
\end{figure}

The loss curves for the quantum models and the hinge-loss diagnostic for the quantum support vector machine appear in Fig.~\ref{fig:loss-quantum}, and the ROC curves and F1-scores in Figs.~\ref{fig:roc-quantum} and~\ref{fig:f1-quantum}. Even with sixteen principal components and only eight qubits, the two trainable embeddings reach an area under the curve near 0.97 and a recall near 0.95, while the recurrent and kernel variants stay well below. The confusion matrix of the quantum convolutional network in Fig.~\ref{fig:cm-quantum} shows high signal efficiency with controlled misclassification, whereas the quantum long short-term memory network and the quantum support vector machine accumulate many misclassified events, which points to limited separability under the imposed constraints.

\section{Discussion}\label{sec:discussion}

\subsection{The strongest models}\label{sec:best}

The lead of the artificial neural network fits the structure of the task. The label is set by an invariant-mass window, and the inputs carry the kinematic components from which the invariant mass is reconstructed, so an expressive feed-forward network can approximate the implicit mass-window boundary smoothly while also using correlations in $\Delta R$, $\Delta\eta$ and $p_{T,\mathrm{tot}}$. Its high precision and recall show that it handles the class balance without sacrificing signal efficiency. Among the quantum models, the quantum convolutional and quantum neural networks reach an accuracy near 91 percent and an area near 0.97. Both are trainable hybrids in which the quantum circuit provides a nonlinear embedding and a classical head performs the final discrimination. The edge of the convolutional variant is consistent with the inductive bias of its convolution-and-pooling structure, which supports a hierarchical extraction of features even within the eight-qubit limit. That both reach recall above 0.95 matters for a trigger, where losing true signal is costly.

\subsection{Why the convolutional and recurrent networks trail the feed-forward one}\label{sec:cnn-lstm}

The convolutional and recurrent networks land near 90 percent accuracy, a little short of the feed-forward network. For the convolutional network, the convolutional bias is not well matched to tabular kinematics, since the features are neither ordered nor embedded in a spatial locality, so the advantage a convolution usually brings is muted. For the recurrent network, the inputs carry little genuine sequential structure; the ordering of the features is essentially artificial, so the gates act more as an alternative parameterization of a feed-forward mapping than as a model of temporal dependence. This explains why its performance is close to the convolutional network but does not exceed the feed-forward one.

\subsection{The support vector machine and its quantum counterpart at a fixed operating point}\label{sec:svm-qsvm}

The classical support vector machine sits below the neural models and would likely benefit from wider hyper-parameter tuning and a better kernel. Its relatively high recall paired with a lower precision means that its boundary accepts many positives at the cost of false positives, which is not the trade-off one wants when the trigger budget is tight. The quantum support vector machine is the weakest model in the benchmark. Its kernel is fixed rather than trained end to end, the feature map is not adapted to the data, and kernel construction scales quadratically with the number of training events, which forces subsetting and can hurt generalization when the full test set is evaluated. A fixed kernel does not capture the fine structure of the decision boundary as well as a trainable hybrid.

\subsection{Why the recurrent quantum model falls behind the trainable embeddings}\label{sec:qlstm}

The quantum long short-term memory network falls well below the quantum convolutional and quantum neural networks. The likely reason is that the sequence it consumes is built from principal components rather than from a physically ordered time series, so its recurrent assumptions do not match the data. On top of that, the repeated circuit evaluations inside the recurrent loop make optimization harder within a fixed epoch budget and under simulation constraints, which shows up as a lower and less consistent area under the curve.

\subsection{Limitations}\label{sec:limits}

Several limitations bound how far the numbers can be pushed, and stating them plainly matters more than the ranking itself.

The most important one concerns the label. Because the window is defined on the invariant mass and the inputs are the four-vector components used to compute that mass, the models are in large part reconstructing a known analytic function from its own ingredients. A high area under the curve is therefore expected and should not be read as discovery of nontrivial structure. A cleaner design would withhold the exact components that enter the mass, or define the label through a physically independent criterion, and would quantify how much of the performance follows directly from the label definition. An explicit analytic baseline that cuts on the reconstructed mass would set the ceiling against which every model should be judged.

A second limitation is that the classical and quantum families do not see the same inputs. The classical models use the full feature set, while the quantum models use sixteen principal components, so the comparison mixes the effect of the paradigm with the effect of the representation. A resource-matched setting, in which the classical models also receive the sixteen components or the quantum models receive comparable information, would separate the two.

A third limitation is statistical. The results come from a single split and a single seed, without error bars, cross-validation or repeated runs, so a gap such as the one between the quantum neural network and the quantum convolutional network can fall within the noise. A benchmark that proposes a ranking needs a mean and a spread over several seeds and a test of significance for the close cases.

A fourth limitation concerns provenance and reproducibility. The dataset is described briefly and cited through a public mirror rather than through the CMS open-data record with its persistent identifier. A precise account of the object definitions, which the window near the $J/\psi$ mass suggests are muon pairs, of the selection flow and of the preprocessing, together with released code and seeds, would make the study reproducible.

\subsection{Implications for triggering and future work}\label{sec:trigger}

From the point of view of a trigger, the operational quantities are the signal efficiency and the background acceptance. Accuracy is the metric reported here, but the high recall of the artificial neural network and of the strongest quantum models indicates that they can reach high signal efficiency at a useful purity, which is what a resonance-like selection needs. The high areas under the curve, at or above 0.96 for the feed-forward, convolutional and recurrent networks and at or above 0.97 for the two trainable quantum embeddings, indicate good score ordering, so a smooth trade-off between rate and efficiency should be reachable by moving the threshold away from the reported 0.5.

The natural next steps follow from the limitations. A deployment-grade study should add an explicit analytic mass-cut baseline, report rate against efficiency and turn-on curves as a function of variables such as $p_{T,\mathrm{tot}}$, measure inference latency and throughput for both families, repeat the training over several seeds with error bars, match the input representation across paradigms, and select the operating point from a bandwidth budget rather than from a fixed threshold. Encoding and ansatz studies for the quantum models, together with a scan of the principal-component dimension, would also clarify why the recurrent and kernel variants fall behind.

\section{Conclusion}\label{sec:conclusion}

Under a shared protocol and a set of physics-motivated features, the classical support vector machine is the weakest of the eight models, while the artificial neural network is the strongest, at 93.53 percent accuracy and 0.9819 ROC-AUC, with the convolutional and recurrent networks close behind. Among the quantum models the trainable hybrid embeddings are the most suitable for this task, with the quantum convolutional and quantum neural networks reaching roughly 90.6 to 90.9 percent accuracy and areas near 0.97, while the fixed-kernel and recurrent quantum variants trail under the eight-qubit and sixteen-component budget. Read honestly, the study shows parity rather than advantage for quantum models in this regime, and its value lies in the controlled reference numbers and in the explicit account of what still needs to be done: rate-constrained operating points, turn-on and efficiency studies over the quantum parameter space, and measurements of latency and robustness relevant to deployment.

\section*{Acknowledgements}
AR acknowledges CIC-UMSNH (Mexico) under grant 18371.
\bibliography{references}

\end{document}